\documentclass[twocolumn,preprintnumbers,amsmath,superscriptaddress,amssymb,aps] {revtex4}
\usepackage{graphicx}
\usepackage{dcolumn}
\usepackage{bm}

\usepackage{float}

\begin{document}
\title {Strain and onsite-correlation tunable quantum anomalous Hall phases in
ferromagnetic (111) LaXO$_3$ bilayers (X$=$Pd, Pt)}
\author{Hai-Shuang Lu}
\address{Department of Physics and Center for Theoretical Sciences, National Taiwan University,Taipei 10617, Taiwan}
\address{College of Physics and Electronic Engineering, Changshu Institute of Technology, Changshu 215500, P. R. China}
\author{Guang-Yu Guo}
\email{gyguo@phys.ntu.edu.tw}
\address{Department of Physics and Center for Theoretical Sciences, National Taiwan University,Taipei 10617, Taiwan}
\address{Physics Division, National Center for Theoretical Sciences, Hsinchu 30013, Taiwan}
\date{\today}

\begin{abstract}
Quantum anomalous Hall (QAH) phases in magnetic topological insulators are characterized by
the scattering-free chiral edge currents protected by their nontrivial bulk band topology.
To fully explore these intriguing phenomena and application of topological insulators,
high temperature material realization of QAH phases is crucial.
In this paper, based on extensive first-principles density functional theory
calculations, we predict that perovskite bilayers (LaXO$_3$)$_{2}$ (X = Pd, Pt) imbedded in
the (111) (LaXO$_3$)$_{2}$/(LaAlO$_3$)$_{10}$ superlattices are high Curie temperature ferromagnets 
that host both QAH and Dirac semimetal phases, depending on the biaxial strain and onsite 
electron correlation. In particular, both the direction (the sign of Chern number) 
and spin-polarization of the chiral edge currents are tunable by either onsite electron correlation
or biaxial in-plane strain. Furthermore, the nontrivial band gap can 
be enhanced up to 92 meV in the LaPdO$_3$ bilayer by the compressive in-plane strain, and
can go up to as large as 242 meV when the Pd atoms are replaced by the heavier Pt atoms.
Finally, the microscopic mechanisms of the ferromagnetic coupling and other interesting
properties of the bilayers are uncovered by analyzing their underlying electronic band structures. 
\end{abstract}

\maketitle
\section{Introduction}
In the past decade, various topological insulators~\cite{Hasan10,Qi11,weng} 
have attracted enormous attention because of their fascinating 
transport properties. In particular, transport currents along the gapless 
edge modes on the surface or at the interface between two topologically 
different insulators are unidirectional and robust against scattering
from disorder due to topologically nontrivial properties of their
bulk band structures. The quantum anomalous Hall (QAH) phase, first proposed
by Haldane in his Nobel Prize-winning paper~\cite{Haldane}, is a two-dimensional (2D) bulk ferromagnetic (FM)
topological insulator (Chern insulator) with a nonzero topological invariant known as the
Chern number in the presence of spin-orbit coupling (SOC) but in the absence
of applied magnetic fields~\cite{weng}. Its associated chiral edge modes 
carry dissipationless unidirectional electrical current.
Due to the intriguing nontrivial topological properties and fascinating 
potential application for designing low energy consumption electronics and 
spintronics, extensive theoretical studies have been made recently 
to search for real QAH insulators~\cite{weng}.  Indeed, specific 
material systems such as FM quantum wells~\cite{Liu08}, FM topological
insulator (TI) films~\cite{ryu}, graphene on magnetic substrates~\cite{Qia10,Che11}, 
and noncoplanar antiferromagnetic (AF) layered oxide~\cite{Zhou16} have been predicted.

Excitingly, this intriguing QAH phase was recently observed in the 
Cr-doped (Bi,Sb)$_2$Te$_3$ films\cite{xue}. However, the QAH phase 
occurs only at very low tempetures due to the small band gap, 
weak magnetic coupling and low carrier mobility in the sample. 
This hampers further exploration of the novel properties of Chern insulators and
also their applications. The low carrier mobility could result from
the disorder due to the doped magnetic impurities in the sample,
while the weak magnetic coupling could originate from the localized Cr
3$d$ orbitals which hardly overlap with the orbitals of the neighboring 
Cr atoms. The problems of the weak magnetic coupling and small band gap 
could be overcomed by introducing 4$d$ and 5$d$ transition metal atoms 
which simultaneously have more extended $d$ orbitals and stronger SOC. 
Clearly, it would be fruitful to search for high temperature QAH phase in
stoichiometry FM 4$d$ and 5$d$ transition metal compounds. 

\begin{figure}
\centering
\includegraphics[width=0.4\textwidth]{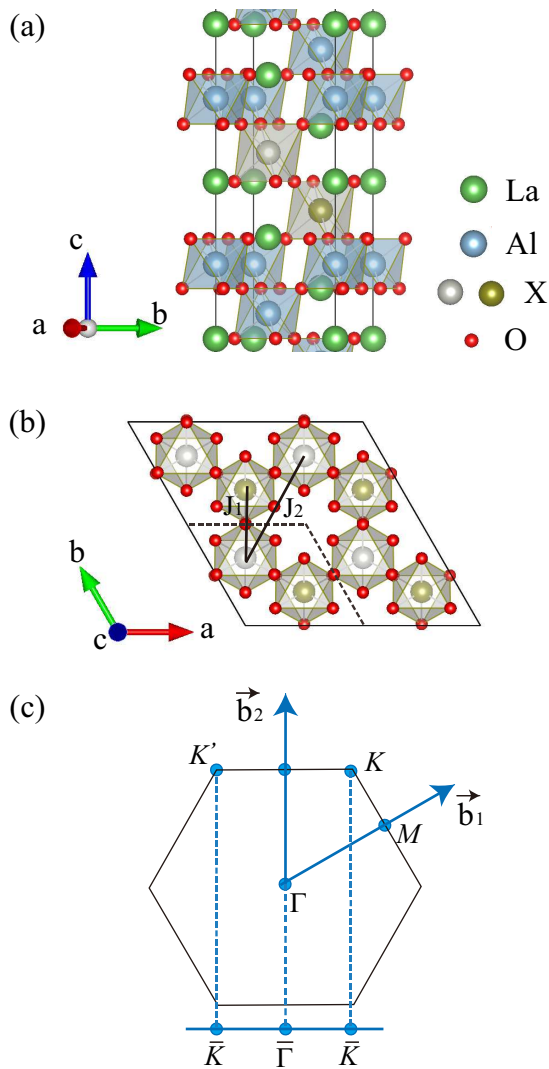}
\caption{\label{fig:Fig1} Crystal structure of a
(LaXO$_3$)$_{2}$/(LaAlO$_3$)$_{10}$ superlattice. (a) Side view of
the unit cell containing the (LaXO$_3$)$_{2}$ bilayer, and (b) Top view
of the (LaXO$_3$)$_{2}$ bilayer forming a buckled honeycomb lattice. 
Two different colors denote the X atoms on the two different planes 
in the bilayer. In (b), black dotted lines indicate the lateral 
primitive unit cell. (c) The Brillouin zone of the honeycomb lattice 
with the reciprocal lattice vectors $\vec{b}_{1}$ and $\vec{b}_{2}$.
$\Gamma$(0,0,0), $K$(2$\pi$/3a,2$\pi$/3a,0), and $M$($\pi$/a,0,0) 
are the high-symmetry points in the Brillouin zone. 
Also in (c), $\bar{\Gamma}$ and $\bar{K}$ are the high symmetry
points in the one-dimensional Brillouin zone for the boundary (edge)
along $\vec{a}$ (1,0,0) direction.}
\end{figure}

\begin{figure}
\centering
\includegraphics[width=0.4\textwidth]{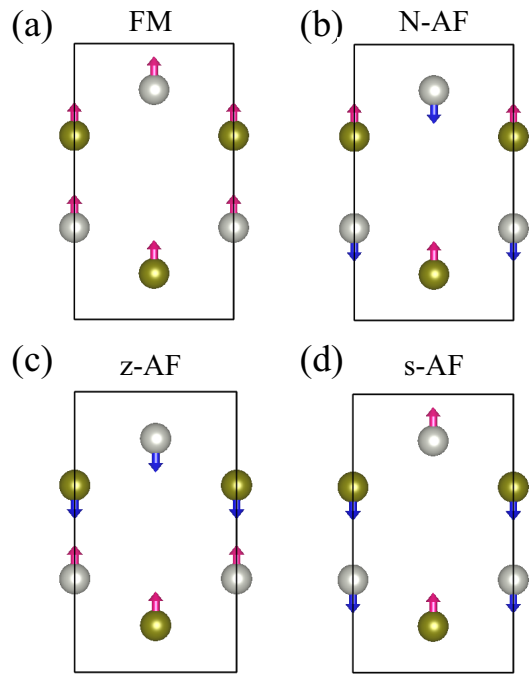}
\caption{\label{fig:Fig2} (a) Ferromagnetic (FM), (b)
N$\acute{e}$el-antiferromagnetic (AF) (N-AF), (c) zigzag-AF (z-AF) and 
(d) stripy-AF (s-AF) configurations in the (LaXO)$_2$/(LAO)$_{10}$ 
superlattices. Arrows represent the directions of spin moments on the X atoms. }
\end{figure}

Transition metal oxides (TMOs) ocver a wide range of crystalline
structures and exhibit a rich variety of fascinating properties
such as charge-orbital ordering, high temperature
superconductivity, colossal magnetoresistance and half-metallic
behavior~\cite{Kobayashi}. Artificial atomic scale TMO
heterostructures offer the prospect of further enhancing these
fascinating properties or of combining them to realize novel
properties and functionalities~\cite{Mannhart,Hwang12} such as the
conductive interface between two insulating
oxides~\cite{Hwang04,Lee16,Lu15}. Recently, based on their tight-binding (TB) 
modelling and first-principles density functional theory (DFT) calculations, Xiao {\it et al.} proposed
in their seminal paper~\cite{Xiao11} that various quantum topological phases
could be found in a class of (111) TMO perovskite bilayers sandwiched
by insulating perovskites where the transition metal atoms in the bilayers
form a buckled honeycomb lattice which is known to host such topological 
phases as QAH phase~\cite{Haldane}. Subsequently, the electronic
structure of a large number of (111) TMO perovskite bilayers
and also double-perovskite monolayers in (111) oxide superlattices
were investigated and some of them were indeed predicted to host the 
quantum spin Hall (QSH), QAH and other topological phases
(see Refs. [\onlinecite{weng}], [\onlinecite{Chandra17}] 
and [\onlinecite{Xiao18}] as well as references therein).

In addition, artificial TMO heterostructures nowadays could be prepared
with atomic precision, thus providing considerable tunability over
fundamental parameters such as SOC, strain and electron correlation.
By varying these fundametal parameters, one could then engineer and also 
manipulate a number of interesting phenomena such as charge, spin and 
orbital orderings, metal-insulator transitions, multiferroics and 
superconductivity in the TMO heterostructures.~\cite{Hwang12}
For example, fascinating phases such as spin-nematic, Dirac half-metallic,
QSH, QAH and fully polarized netamtic phases could emerge in (111) (LaNiO$_3$)$_{2}$ 
bilayer as the strength of the onsite Coulomb repulsion ($U$) is varied.~\cite{AR2011,Yangky}
Moreover, based on his low-energy effective models calculations,
Okamoto~\cite{Okamoto} recently proposed that (111) 4$d$ and 5$d$ TMO 
perovskite bilayers such as (111) (LaPdO$_3$)$_{2}$ bilayer 
where the effects of the SOC and electron correlation are comparable, 
would provide unique playgrounds for studying quantum phases 
caused by the interplay of the SOC and electron correlation.
Lattice strains can also significantly influence the band structure
and electronic properties~\cite{Lee16,Liu14,Wang16}, and are thus 
another powerful parameter for tuning material properties. 
Indeed, Liu {\it et al.}\cite{Liu14} recently showed that 
topological phase transitions in narrow-gap 
semiconductors could be engineered by applying strains.

In this paper, we consider (111) bilayers (LaXO$_3$)$_{2}$ 
of 4$d$ and 5$d$ transition metal (X) perovskites (X = Pd, Pt) 
embeded in an insulating perovskite LaAlO$_3$ matrix. 
In the (111) (LaNiO$_3$)$_{2}$ bilayer, although the strong 
onsite correlation could drive it into the QAH phase~\cite{AR2011,Yangky}, 
it remains in the ferromagnetic metallic phase in the realistic $U$ 
value range~\cite{AR2012} perhaps because of the small SOC on the Ni atoms. 
Therefore, it would be interesting to see how the electronic properties
especially the topology of the band structure would change 
when the Ni atoms are replaced by the isoelectronic Pd or Pt atoms 
which have stronger SOCs. 
Furthermore, perovskite LaPdO$_3$ ~\cite{Kimseungjoo,
Kimseungjoo2} has been synthesized and the Pd ion was found to have the same electronic
configuration ($t_{2g}^{6}e_{g}^{1}$) as the Ni ion in perovskite LaNiO$_3$, thus
making the fabrication of its (111) heterostructures feasible.

Therefore, here we perform a systematic first-principles DFT
study on the magnetic and electronic properties
of these (111) (LaXO$_3$)$_{2}$ bilayers under various biaxial strains
as a function of onsite electron correlation. 
First of all, we find that both bilayers (LaXO$_3$)$_2$ (X = Pd, Pt) 
are ferromagnetic with high Curie temperatures. Furthermore,
with weak onsite Coulomb repulsion, both systems are 
Dirac semimetals with Dirac points located slightly above and
below the Fermi level. Topologically nontrivial gaps are opened
at these Dirac points when the SOC is included. Secondly, 
bilayer (LaPdO$_3$)$_2$ [(LaPtO$_3$)$_2$] becomes a Chern insulator
with Chern number $C = +1$ when the $U$ value on Pd [Pt] is increased
to 3.5 eV [2.0] eV which bring the Dirac points to the Fermi level. 
Remarkably, bilayer (LaPdO$_3$)$_2$ [(LaPtO$_3$)$_2$] becomes a different
Chern insulator with Chern number $C = -1$ when the $U$ value 
on Pd [Pt] is further increased slightly to 3.6 eV [2.5] eV. 
Interestingly, this shows that the direction of chiral edge current 
could be switched by tuning the $U$ value.
Thirdly, we uncover that the biaxial strain could also drive 
bilayers into topological phase transitions and also enlarge
the topological band gaps as well as lower the critical $U$ values
for the metal-insulator transitions. For example, the critical $U$ value 
for the metal-insulator transition is reduced to 2.8 eV for bilayer (LaPdO$_3$)2 
under -3 \% strain, and to 1.5 eV for bilayer (LaPtO$_3$)2 under -3.5 \% strain.
Finally the topological band gap
in bilayer (LaPtO$_3$)$_2$ could reach as large as 150 meV
which is well above room temperature.
All these interesting findings thus demonstrate that bilayer (LaPdO$_3$)$_2$ 
and (LaPtO$_3$)$_2$ are valuable 
quasi-2D materials for exploring such novel electronic
phases as QAH effect at high temperatures and also 
for technological applications such as low-power consumption
nanoelectronics and oxide spintronics.

\section{Structures and methods}
We consider perovskite bilayers (LaXO$_3$)$_{2}$
sandwiched by an insulating perovskite LaAlO$_3$ slab as in the
(LaXO$_3$)$_{2}$/(LaAlO$_3$)$_{10}$ superlattices grown along the
[111] direction, where X is either 4$d$ transition metal Pd or 
5$d$ transition metal Pt. The resultant superlattices have a 
trigonal symmetry ($D_{3d}$), and the X atoms in each bilayer form 
a buckled honeycomb lattice (see Fig. 1).
Since the LaAlO$_3$ slab is much thicker than the
(LaXO$_3$)$_{2}$ bilayer, the LaAlO$_3$ slab could be regarded as
the substrate. Therefore, we fix the in-plane lattice constant ($a$)
of the superlattices to $a=\sqrt{2}a_0$, and $a_0$
is the theoretical lattice constant of bulk LaAlO$_3$
perovskite. The calculated $a_0$ is 3.81 \AA{}, being close to the experimental value of
3.79 \AA\cite{Geller}. With lattice constant $a$ and 
symmetry $D_{3d}$ fixed, the lattice constant $c$ and the internal coordinates of
all the atoms are theoretically optimized.
Note that within this $D_{3d}$ symmetry constraint, the metal
atoms in these superlattices could relax only in the
[111]-direction (i.e., out-of-plane direction), 
although the oxygen atoms could move both laterally and vertically.

The electronic and magnetic structure are calculated
based on the DFT with the generalized
gradient approximation (GGA) in the form of Perdew-Berke-Ernzerhof
\cite{perdew}. The accurate projector augmented wave (PAW)
method\cite{paw1}, as implemented in the VASP code\cite{kresse1,kresse2},
is uesd. The fully relativistic PAW potentials are adopted in
order to include the SOC. The valence configurations of La, Al,
Pd, Pt and O atoms used in the calculations are
4$p$$^{6}$5$s$$^{2}$6$d$$^{1}$, 3$s$$^{2}$3$p$$^{1}$,
4$p$$^{6}$4$d$$^{9}$5$s$$^{1}$, 5$p$$^{6}$5$d$$^{9}$6$s$$^{1}$,
2$s$$^{2}$2$p$$^{4}$, respectively.
A plane wave cutoff energy of 450 eV and the total energy convergence criteria 
of 10$^{-5}$ eV are used throughout. A fine Monkhorst-Pack
$k$-mesh of $20\times 20\times 4$ is used in the selfconsistent calculations.

Because the $d$-orbitals of the 4$d$ and 5$d$ transition metal
elements are rather extended, the onsite electron-electron repulsion is
expected to be moderate but is nonetheless comparable to the
strong SOC in these bilayer systems. The interplay of the
comparable onsite electron correlation and SOC can lead
to such fascinating effects as wide gap Chern Mott insulating
phases in (111) 4$d$ and 5$d$ transition metal perovskite bilayers
\cite{Guo17} as well as metal-insulator transition, strong
topological insulating phase and quantum spin liquid phase in
pyrochlore iridates \cite{Car12,Car13,Che15,Sch16}. Therefore, 
in the present calculations, the onsite Coulomb repulsion ($U$) 
on the Pd and Pt atoms is also taken into account within the 
GGA + $U$ scheme~\cite{Dud98}. The onsite Coulomb repulsion $U$ 
is varied between 0 and 4 eV for Pd and between 0 and 3 eV for Pt. 

To understand the magnetic interactions and also to estimate the
ferromagnetic ordering temperature ($T_C$) in the bilayers, 
we consider all possible magnetic configurations in the supercell 
containing two chemical formula units (see Fig. 2), namely, the FM,
N$\acute{e}$el-antiferromagnetic (N-AF), zigzag-antiferromagnetic
(z-AF) and stripy-antiferromagnetic (s-AF) structures. To 
evaluate the first-neighbor ($J_1$) and second-neighbor ($J_2$)
exchange coupling parameters [see Fig. 1(b)], the calculated total 
energies of the FM, N-AF and z-AF magnetic configurations are mapped 
to the classical Heisenberg model $H = E_{0} -
\sum_{i>j}J_{ij}({\hat{e}_{i}}.{\hat{e}_{j}})$ where $J_{ij}$ is
the exchange coupling parameter between sites $i$ and $j$, and 
${\hat{e}_{i(j)}}$ denotes the direction of spin on site $i(j)$.
This results in $J_1 = (E_{N-AF}-E_{FM})/12$ and $J_2 = (E_{z-AF}-E_{FM}-4J_1)/12$.

The anomalous Hall conductivity (AHC) for all the ferromagnetic
superlattices is calculated based on the Berry-phase
formalism\cite{Xiao10}. Within this Berry-phase formalism, the AHC
($\sigma_{ij}^A$ = $J^{c}_{i}$/$E_{j}$) is given as a BZ integration
of the Berry curvature for all the valence bands,
\begin{equation}
\sigma_{xy}^A = -\frac{e^{2}}{\hbar}\sum
\limits_{n\in{VB}}\int_{BZ}\frac{d{\bf k}}{(2\pi)^{3}}{\Omega^{n}_{xy}}({\bf k}),
\end{equation}
\begin{equation}
{\Omega^{n}_{xy}}({\bf k}) = -\sum
\limits_{n'\neq{n}}\frac{2Im[p^{x}_{ij}p^{y}_{ji}]}{(\epsilon_{{\bf k}n}-\epsilon_{{\bf k}n'})^{2}}
\end{equation}
where $\Omega$$^{n}_{ij}({\bf k})$ is the Berry curvature for the $n$th
band at ${\bf k}$. $J^{c}_{i}$ is the $i$-component of the charge current
density $\textbf{J}^{c}$ and $E_{j}$ is the $j$-component of the
electric field $\textbf{E}$. Since a large number of $k$-points
are needed to get accurate AHCs, we use the efficient Wannier function
interpolation method\cite{x_wang,lopez} based on maximally
localized Wannier functions (MLWFs) \cite{Marzari}. Since the
energy bands around the Fermi level are dominated by X
$e_{g}$-orbitals, eight MLWFs per bilayer of X $e_{g}$
($d^{\uparrow},d^{\downarrow}$) orbitals are constructed by
fitting to the relativistic band structure in the energy window from
-1.5 eV to 1.7 eV for the (LaPdO$_3$)$_2$ bilayer and from -1.5 eV
to 0.6 eV for the (LaPtO$_3$)$_2$ bilayer. The band structure
obtained by the Wannier interpolation agrees well with that from
the DFT calculation. The AHC ($\sigma_{xy}^A$) for both
bilayers was then evaluated by taking a very dense $k$-point mesh
of $300\times{}300{}\times10$ in the Brillouin zone.

\section{Results and discussion}
\begin{table}
\caption{The total energy ($E$) (relative to that of the FM
state) and total spin magnetic moment ($m_{s}^{t}$) as well as atomic
spin magnetic moments of the X atoms ($m_{s}^{X}$) and the O atoms ($m_{s}^{O}$)
that connect two adjacent X atoms in the central layer [Fig. 1(a)]
of the converged magnetic configurations (Fig. 2) in the 
(LaXO$_3$)$_{2}$ bilayers calculated with $U = 4$ eV for Pd and $U = 3$ eV for Pt.
Note that the initial N-AF magnetic structure always relaxes to the
NM state and thus is not listed here. Also, the initial s-AF magnetic
configuration converges to the s-FM state, as listed below.
}
\begin{tabular}{ccccccccccc}
\hline\hline
                &                        & FM         & NM        & z-AF         & s-FM  \\
\hline
(LaPdO$_3$)$_2$ &$E$ (meV/cell)          & 0.0        & 355.8     & 106.2        & 86.2  \\
           &$m_{s}^{t}$ ($\mu_{B}$/cell) & 2.0        & 0.0       & 0.0          & 0.98   \\
           &$m_{s}^{Pd}$ ($\mu_{B}$/atom)& 0.81       & 0.0       & 0.70         & 0.79  \\
                &                        & 0.81       & 0.0       & -0.70        & 0.04  \\
           &$m_{s}^{O}$ ($\mu_{B}$/atom) & -0.034     & 0.0       & -0.044       & -0.024 \\
                &                        & -0.034     & 0.0       &  0.044       & -0.024 \\
                \hline
(LaPtO$_3$)$_2$ &$E$ (meV/cell)          & 0.0        & 337.0     & 103.7        & 83.4  \\
                &$m_{s}^{t}$ ($\mu_{B}$) & 2.0        & 0.0       & 0.0          & 1.0\\
           &$m_{s}^{Pd}$ ($\mu_{B}$/atom)& 0.83       & 0.0       & 0.69         & 0.78   \\
                &                        & 0.83       & 0.0       & -0.69        & 0.06   \\
           &$m_{s}^{O}$ ($\mu_{B}$/atom) & -0.015     & 0.0       & -0.022       & -0.013 \\
                &                        & -0.015     & 0.0       &  0.022       & -0.013 \\
\hline\hline
\end{tabular}
\label{TabI}
\end{table}

\subsection{Magnetic structure}
As mentioned above, we perform selfconsistent spin-polarized
band structure calculations for the FM, N-AF, z-AF and s-AF 
magnetic configurations (Fig. 2) in the (LaXO$_{3}$)$_{2}$ bilayers.
Nonetheless, we find that the initial N-AF magnetic configuration
always converges to the nonmagnetic (NM) state, i.e., 
the metastable N-AF magnetic configuration does not exist in these systems.
Interestingly, the initial s-AF magnetic configuration relaxes 
to a stripy-ferromagnetic (s-FM) configuration with different
magnetic moments on the X atoms (Table I). The principal 
properties of all the converged magnetic configurations obtained
with $U = 4$ eV for Pd and $U = 3$ eV for Pt are listed in Table I,
as examples. As Table I shows, the FM state has the lowest
total energy and hence is the ground state magnetic structure.
The s-FM configuration has a higher total energy
than the FM state but is lower in total energy than the z-AF configuration.
The NM state has the largest total energy.
For $U = 4$ $(3)$ eV, the calculated magnetic moment on Pd (Pt)
in the (LaPdO$_{3}$)$_{2}$ [(LaPtO$_{3}$)$_{2}$] bilayer
is 0.81 (0.83) $\mu_B$/atom in the FM state, $\pm$ 0.70 ($\pm$ 0.69) $\mu_B$
in the z-AF state, and 0.79/0.04 (0.78/0.06) $\mu_B$ in
the s-FM state (see Table I). Interestingly, the magnetic
moment of the O atom connecting the two adjacent Pd (Pt) atoms 
has a spin polarization being opposite to that of Pd (Pt) in the FM
configuration. For example, the O magnetic moment is -0.034 $\mu_B$ 
and the Pd magnetic moment is 0.81 $\mu_B$ in the (LaPdO$_{3}$)$_2$ bilayer 
at $U = 4$ eV. The opposite spin polarizations of the O and Pd (Pt) atoms
is due to the hybridization of the Pd (Pt) $d$ and O 
$p$ orbitals in the FM state.

\begin{table}
\caption{The exchange coupling parameters ($J_1$ and $J_2$) (see Fig. 1),
X magnetic moment in the FM state, Curie temperature ($T_C$),
band gap ($E_{g}$) and Chern number ($C$) of bilayers (LaXO$_3$)$_{2}$ 
calculated using different $U$ values. 
}
\begin{tabular}{ccccccccccc}
\hline\hline
(LaPdO$_3$)$_2$ &$U$ (eV)           & 0.0    & 2.0     & 3.0     & 3.5      & 4.0\\
                &$J_{1}$ (meV)    & 3.8    & 14.0    & 22.3    & 28.9     &35.3 \\
                &$J_{2}$ (meV)    & 2.57   & 2.67    & 3.76    & 3.94     &4.46 \\
            &$m_s^{Pd}$ ($\mu_B$) & 0.35   & 0.61    & 0.74    & 0.78     & 0.81 \\
                &$T_{C}$ (K)      & 104    & 224     & 346     & 426      &513 \\
                &$E_{g}$ (meV)    & 0      & 0       & 0        & 4        & 50  \\
                &  $C$            & -      & -       & -        & 1       & -1   \\
                \hline
(LaPtO$_3$)$_2$ &$U$ (eV)           & 0.0    & 1.0     & 2.0      & 2.3     & 3.0 \\
                &$J_{1}$ (meV)    & 7.1    & 17.5    & 32.0     & 37.6    & 50.7   \\
                &$J_{2}$ (meV)    & 2.63   & 0.87    & 0.59     & 0.30    & 0.29    \\
            &$m_s^{Pt}$ ($\mu_B$) & 0.45   & 0.60    & 0.75     & 0.78    & 0.83 \\
                &$T_{C}$ (K)      & 144    & 223     & 385      & 443     & 595     \\
                &$E_{g}$ (meV)    & 0      & 0       & 8        & 16      & 150 \\
                &  $C$            & -      & -       & 1        & 1       & -1 \\
\hline\hline
\end{tabular}
\label{TabII}
\end{table}

As described in the preceeding section, the exchange coupling parameters
$J_1$ and $J_2$ could be evaluated once the total energies of the FM, N-AF 
and z-AF magnetic configurations are known. However, as mentioned above,
the metastable N-AF state does not exist in these bilayer systems.
Therefore, we perform further constrained DFT calculations
for the N-AF magnetic configuration for both bilayer structures. 
Using the total energies from these constrained DFT calculations,
we derive the nearest-neighbor and second-neighbor
magnetic coupling parameters between the X atoms in the (LaXO$_{3}$)$_2$
bilayers,  as listed in Table II. All the
calculated exchange coupling parameters are positive and thus the
magnetic interaction between the X atoms is ferromagnetic. In both
systems, $J_{1}$ is larger than $J_{2}$ and increases
rapidly with the $U$ value. In the present Heisenberg model (see Sec. II),
the square of the local magnetic moment size has been incorporated into
the exchange coupling parameters ($J$s). Thus, the rapid increase 
in the $J_{1}$ value with $U$, largely reflects the enhanced X magnetic moment 
due to the strongler onsite Coulomb repulsion $U$ (see Table II),
although the intrinsic first near-neighbor FM coupling
[$J_1/(m_s^X)^2$] does increase gradually as the $U$ increases.
In contrast, $J_{2}$ decreases with the $U$ value in the (LaPtO$_{3}$)$_2$ 
bilayer while it increases slightly in the (LaPdO$_{3}$)$_2$ bilayer.
In the mean-field approximation, $ k_BT_c=\frac {1}{3}\sum z_iJ_i$ 
where $z_i$ is the number of the $J_{i}$ X-X bonds for an X atom. 
Thus, we could roughly estimate magnetic ordering temperature $T_{C}$ 
using the calculated $J_{1}$ and $J_{2}$ values. Table II indicates that 
the estimated Curie temperatures $T_{C}$ increases from $\sim$100 K to 
that ($\sim500$ K) above room temperature as the $U$ increases.
Note that the Curie temperature given by the mean-field approximation 
is generally too high by as large as a factor of 2.~\cite{Feng18}.

\begin{figure}
\centering
\includegraphics[width=0.46\textwidth]{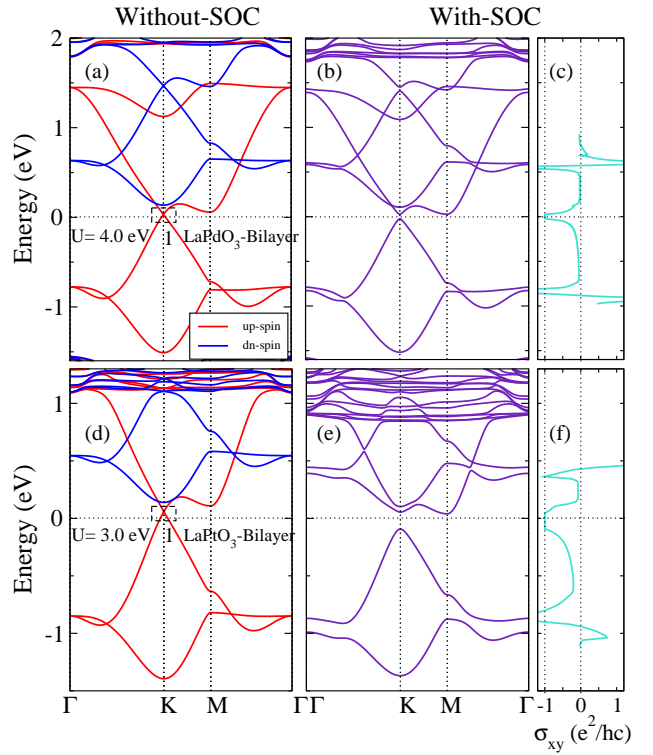}
\caption{\label{fig:Fig3} Band structures without (a,d) and with (b,e)
SOC as well as anomalous Hall conductivity ($\sigma_{xy}$) (c,f) of the
(LaPdO$_{3}$)$_2$ (a,b,c) and (LaPtO$_{3}$)$_2$ (d,e,f) bilayers
for $U=4$ eV on Pd and $U=3$ eV on Pt. Zero refers to the
Fermi level. }
\end{figure}

\subsection{Electronic band structure}

Let us now examine the main features of the electronic band
structures of the (LaXO$_{3}$)$_2$ bilayers. 
The band structures calculated without and with the SOC for the 
FM bilayers with $U=4$ eV for Pd and $U=3$ eV for Pt
are displayed in Fig. 3. In
the absence of SOC, both bilayers are half-metallic, with
the energy bands in the vicinity of the Fermi level being purely
spin-up Pd $e_{g}$ - O $p$ hybridized antibonding bands [see Figs.
3(a) and 3(d)]. This half-metallicity is consistent with the
integer value of the total spin magnetic moment (2.0 $\mu_B$/cell) (Table I). 
A distinctive set of four bands emerges for each spin channel,
namely, rather flat bottom and top bands connected by two dispersive
bands in between, which cross at the K and K$'$ points [in 
the dotted line box labelled 1 in Figs. 3(a) and 3(d)] and thus
form two Dirac cones just above the Fermi level, being consistent 
with the tight-binding model for a honeycomb lattice with a FM open
$e_{g}$ shell \cite{Xiao11}. 
These Dirac nodal points at the K and K' points are 
protected by the trigonal symmetry. The band structures 
presented in Fig. 3 are similar to that of the LaNiO$_{3}$
bilayer~\cite{AR2011,AR2012,Yangky}.
However, the much stronger SOC on the heavier Pd and Pt atoms
should have significant effects on the band
structures of these 4$d$ and 5$d$ transition metal perovskite
bilayers. Indeed, when the SOC is switched-on, the two spin-up 
crossing bands (red curves) hybridize and the Dirac points 
become gapped, resulting in an insulating gap at the Fermi lever
[Figs. 3(b) and 3(e)]. Both (LaPdO$_{3}$)$_2$ and
(LaPtO$_{3}$)$_2$ bilayers are, therefore, SOC-driven insulators 
with the large band gaps of about 50 meV and 150 meV, respectively. 

To understand the nature of the lower conduction bands 
and upper valence bands in a wide energy range, we plot in
Fig. 4 the atom-decomposed densities of states (DOS) for
bulk LaPdO$_{3}$ and its FM (LaPdO$_{3}$)$_2$ bilayer. Since
the (LaPtO$_{3}$)$_2$ bilayer exhibits similar features of the valence
and conduction bands, here we focus on the electronic structure of
bilayer (LaPdO$_{3}$)$_2$ only. In bulk LaPdO$_{3}$, the
valence bands extending from -7.0 to about -1.2 eV are the strongly 
Pd $t_{2g}$ and O $p$ orbital hybridized bonding bands with the 
Pd $t{_{2g}^{{6}\uparrow\downarrow}}$ shell fully occupied [Fig. 4(a)].
The lower conduction bands ranging from -0.6 eV to 2.8 eV,
on the other hand, consist of the Pd $e_{g}$ and O $p$ orbital
hybridized antibonding bands and are partially filled.
Thus bulk LaPdO$_{3}$ is predicted to be a NM metal.
Experimentally, bulk LaPdO$_{3}$ is found to be a paramagnetic 
metal~\cite{Kimseungjoo,Kimseungjoo2}   
and the formal valence of Pd is $3+$. This corresponds to a partially 
filled $4d$ shell of  $t_{2g}^{6}${e$_{g}^{1}$}, i.e., one
electron in the doubly degenerate $e_{g}$ manifold and six electrons
completely filling the $t_{2g}$ shell. The calculated atom-decomposed 
DOSs displayed in Fig. 4(a) are consistent with these experimental results.

In the (111) (LaPdO$_{3}$)$_2$/(LaPdO$_{3}$)$_{10}$ superlattice, 
the transition metal perovskite bilayer is sandwiched by an wide-band 
gap perovskite LaAlO$_{3}$ slab. Thus, the conduction bands made up of
transition metal $d$ orbitals are confined within the bilayer. 
Consequenttly, in these superlattices,
an important difference is that the eight X $e_{g}$-dominated conduction 
bands near the Fermi level are narrower. For example, the bandwidth of
the spin-down Pd $e_{g}$ conduction band gets significantly
reduced from about 3.2 eV in bulk LaPdO$_{3}$ to just $\sim$2.0 eV 
in the (111) (LaPdO$_{3}$)$_2$/(LaPdO$_{3}$)$_{10}$ superlattice (see Fig. 4). 
This conduction band narrowing results mainly from two effects, 
namely, that the hopping within each bilayer must proceed by a repeated
90-deg change in the hopping direction in the cubic lattice and that the
X atoms in each bilayer undergo a significant reduction of the coordination number. 
The significant narrowing of the conduction band leads to enhanced intra-atomic
exchange interaction among the $e_{g}$ electrons and thus results in the formation
of the Pd local spin magnetic moment. The resultant Pd magnetic moments
are then coupled ferromagnetically via the nonmagnetic O atom that
connects the two neighboring Pd atoms sitting, respectively,
in the upper and lower Pd layers [see Figs. 1(a) and 1(b)].
As pointed out by Kanamori and Terakura~\cite{Kanamori001},
the spin-up O $p$ orbitals would hybridize with the spin-up
$e_g$ orbitals of the neighboring Pd atoms and form the Pd $e_g$ 
and O $p$ orbital antibonding conduction bands. Consequently,
the spin-up O $p$ conduction band would be pushed up slightly
and this would result in an energy gain by transfering some
electrons from the spin-up O $p$ band to the spin-down O $p$ band.
Thus, the nonmagnetic O atom would become negatively spin-polarized
and the small resultant O spin magnetic moment is antiparallel
to that of the neighboring Pd atoms (see Table I). The size
of this induced O spin magnetic moment
would thus reflect the strength of such ferromagnetic coupling
between the neighboring Pd atoms. 
Clearly, this Kanamori and Terakura mechanism would not work
if the neighboring Pd atoms are to couple antiferromagnetically, 
since no induced O magnetization would be possible and hence no energy gain 
would occur in this case.~\cite{Kanamori001}
  
\begin{figure}
\centering
\includegraphics[width=0.4\textwidth]{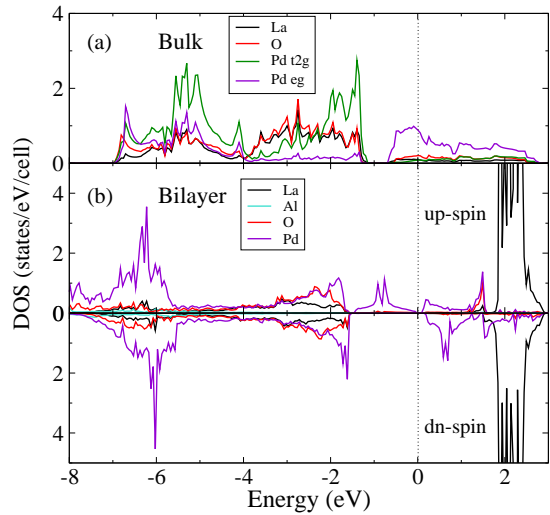}
\caption{\label{fig:Fig4} Atom-decomposed densities of states (DOS)
of bulk LaPdO$_3$ and its FM (LaPdO$_{3}$)$_2$ bilayer calculated
with SOC and $U = 4$ eV.}
\end{figure}

\subsection{Quantum anomalous Hall phases vs. onsite correlation}
As mentioned before, the FM (LaXO$_{3}$)$_2$ bilayers are
found to be semiconductors with the insulating gap opened near
the Dirac points when the SOC is included. We thus could expect
that the band gap would be topologically nontrivial and hence the
bilayers could be Chern insulators. To verify the topological nature of
these insulating gaps, we calculate the AHC ($\sigma_{xy}$) for
these bilayers. For a three-dimensional (3D) quantum Hall
insulator, $\sigma_{xy} = C$ e$^2$/h$c$ where $c$ is the lattice
constant along the $c$-axis normal to the plane of longitudinal
and Hall currents and $C$ is an integer known as the Chern number ~\cite{Hal87}. 
For a normal FM insulator, on the other
hand, $\sigma_{xy} = 0$. The calculated AHC of the (LaXO$_{3}$)$_2$
bilayers is displayed in Fig. 3 for $U = 4$ eV on Pd and $U = 3$
eV on Pt. Indeed, Figs. 3(c) and 3(f) show that in both FM (LaXO$_{3}$)$_2$ 
bilayers, $\sigma_{xy}$ = -1 e$^2$/h$c$ in the gap regions.
This demonstrates that both bilayers are QAH insulators with
Chern number $C = -1$. 
\begin{figure}
\centering
\includegraphics[width=0.47\textwidth]{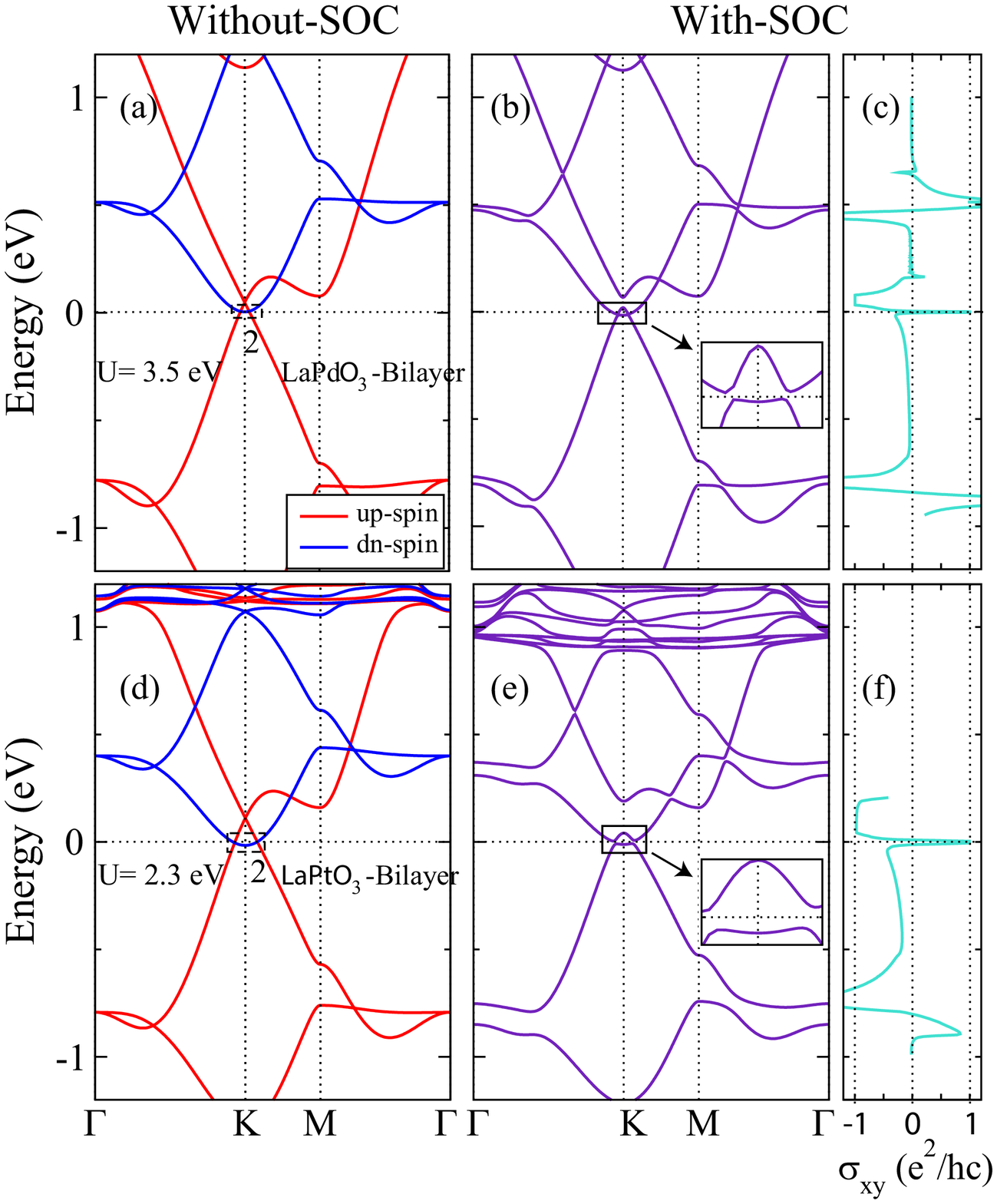}
\caption{\label{fig:Fig5} Band structures without (a,d) and with (b,e)
SOC as well as anomalous Hall conductivity ($\sigma_{xy}$) (c,f) of the
(LaPdO$_{3}$)$_2$ (a,b,c) and (LaPtO$_{3}$)$_2$ (d,e,f) bilayers
for $U=3.5$ eV on Pd and $U=2.3$ eV on Pt. Zero refers to the
Fermi level. }
\end{figure}

To see how the band structure and topological phase change 
when the strength of onsite Coulomb repulsion $U$ is varied, 
we plot in Figure 5 the band structures calculated without and with the
SOC for $U =3.5$ eV on Pd and $U = 2.3$ eV on Pt. Interestingly,
the spin-up and spin-down X $e_{g}$ dominant bands in
the bilayers now cross each other at the Fermi energy, resulting in
that the two bilayers are a metal [Figs. 5(a) and 5(d)].
Compared with the band structure with a larger $U$ value
on Pd and Pt [Figs. 3(a) and 3(d)], the spin-down bands move
slightly downward because of the smaller X magnetic moments
(see Table II) and thus the smaller exchange splitting 
of the spin-up and spin-down $e_{g}$ dominant bands.
When the SOC is switched-on, these crossing points
become gapped with a global band gap of $\sim$4.0 meV 
in the (LaPdO$_{3}$)$_2$ bilayer and of $\sim$16.0 meV
in the (LaPtO$_{3}$)$_2$ bilayer [see the enlarged view 
around the K point in the inset in Figs. Fig. 5 (b) and 5(e), respectively]. 
The band structures of this kind are similar to that of the 
graphene-based heterostructure \cite{zhenhuaqiao14,JZhang2015,ZWang2015}, 
in which a band inversion between two $\pi$ bands 
is induced by the SOC and this band inversion results in
a nontrivial band topology.
In fact, the band gaps between the different spin-polarized 
X $e_{g}$ bands in the (LaXO$_{3}$)$_2$ bilayers here 
[Figs. 5(b) and 5(e)] are also topologically nontrivial.
Indeed, Figs. 5(c) and 5(f) show that the calculated 
$\sigma_{xy}  = +1$ e$^2$/h$c$ within the gap, indicating
that both bilayers are Chern insulators with Chern number $C = +1$. 
Interestingly, the Dirac points of the spin-up bands at the K point
mentioned above, now move above the Fermi level 
[Figs. 5(a) and 5(d)]. They become gapped when the SOC
is included [see Figs. 5(b) and 5(e)]. 
For the sake of clarity, let us call the gaps of this kind 
the local gaps. The global band gaps calculated with the 
SOC included for different $U$ values are listed in Table II.
It should be noted that this SOC-induced global band gap disappears 
when $U$ is less than 3.5 eV in the (LaPdO$_{3}$)$_2$ bilayer 
and 2.0 eV in the (LaPtO$_{3}$)$_2$ bilayer. Nonetheless, the 
nontrivial local band gaps remain open as long as the SOC is tuned on
(see the Appendix).

\begin{figure}
\centering
\includegraphics[width=0.48\textwidth]{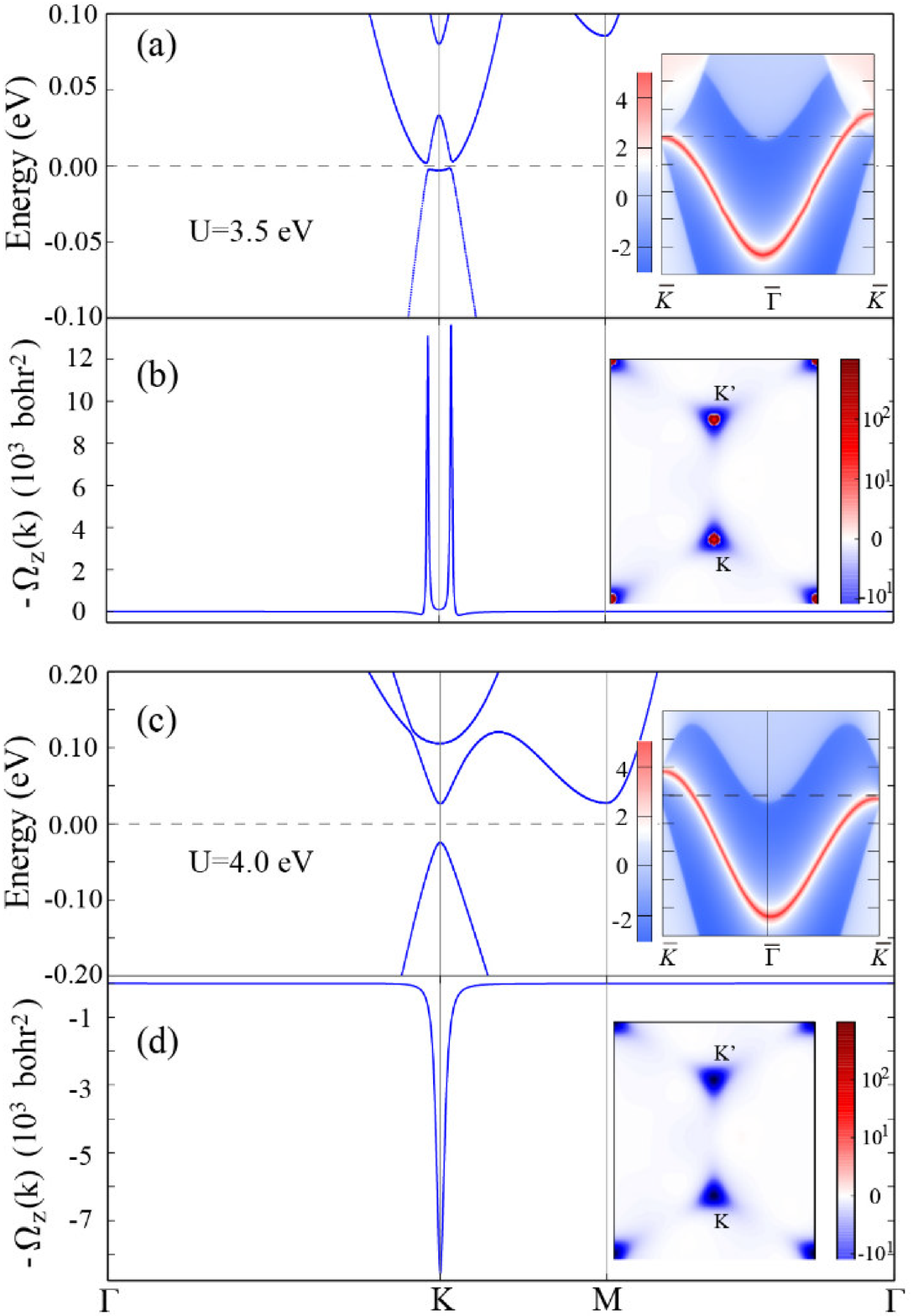}
\caption{\label{fig:Fig6} Energy bands (a,c) and Berry curvature
(b,d) along the symmetry lines in the $k_z = 0$ 2D Brillouin zone for the 
(LaPdO$_{3}$)$_2$ bilayer. The edge band diagrms are given 
in the insets in (a,c), and Berry curvature distributions 
on the $k_z = 0$ plane in the Brillouin zone are displayed 
in the insets in (b,d). 
}
\end{figure}

Interestingly, the sign of $\sigma_{xy}$ (Chern number) changes 
from -1 e$^2$/h$c$ (-1) to +1 e$^2$/h$c$ (+1) when the onsite Coulomb 
repulsion is lowered from 4.0 (3.0) eV to 3.0 (2.3) eV for Pd (Pt) 
in the (LaPdO$_{3}$)$_2$ [(LaPtO$_{3}$)$_2$] bilayer.
This implies that the direction of the chiral edge current 
would be reversed by the change of the onsite correlation strength.
The enlarged energy bands near the Fermi level and gap Berry curvature
along the high symmetry lines are displayed in Fig. 6 for
the (LaPdO$_{3}$)$_2$ bilayer with $U = 3.5$ and 4.0 eV for Pd.
The gap Berry curvature distributions on the $k_z =0$ plane are
also shown in the insets in Figs. 6(b) and 6(d). 
One clearly sees the pronounced peaks in the vicinity of the K points
with different signs for the different $U$ cases.
The calculated edge band diagrams are plotted as spectral functions
in Figs. 6(a) and 6(c). In both cases, there is one gapless edge
band crossing the Fermi level. Furthermore, the Fermi velocities
have opposite signs in these two cases. Therefore, as dictated
by the bulk-edge correspondence theorem, the observed one metallic 
edge state is consistent with Chern number $|C| = 1$ and also the 
chirality of the edge state (i.e., the direction of the edge current) 
conforms with the sign of the Chern number (see Table II).    
This finding would suggest the possibility of not only achieving 
the QAH phase but also designing the flow direction of the dissipationless 
edge current in the FM (LaXO$_{3}$)$_{2}$ bilayers by varying the $U$ value.
\begin{figure}
\centering
\includegraphics[width=0.46\textwidth]{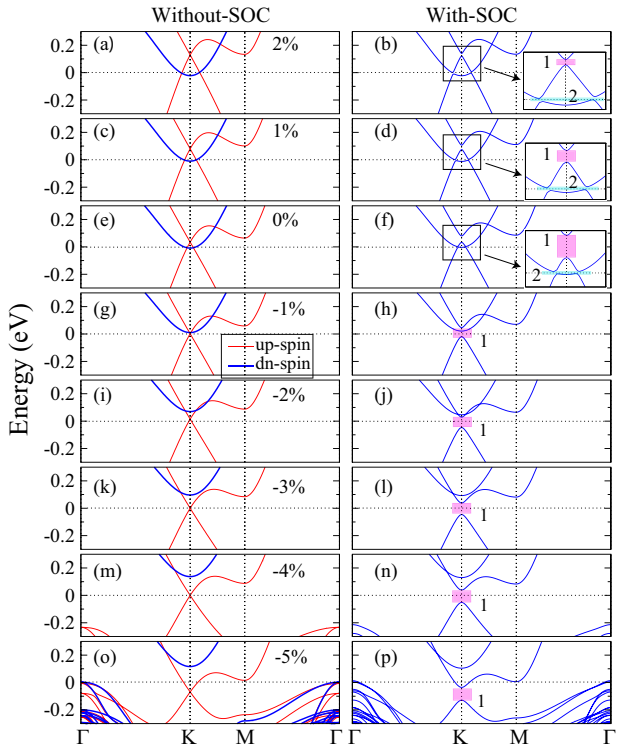}
\caption{\label{fig:Fig7} The band structures calculated without (left
panels) and with (right panels) SOC of the (LaPdO$_{3}$)$_2$
bilayer under various in-plane strains. The $U = 3.5$ eV is
used for Pd. In the left panels,
spin-up and spin-down bands are plotted as red and blue curves,
respectively. In the right panels, the band gap between
two same (opposite) spin bands is labelled as local gap 1 (2).
The Fermi level is at 0 eV.}
\end{figure}

\subsection{Strain-driven topological phase transitions}
Lattice strains can significantly influence the band structure
and electronic properties~\cite{Liu14,Wang16} and are thus an useful parameter
for tuning material properties.
Therefore, we perform systematic DFT calculations for 
the (LaXO$_{3}$)$_2$ bilayers under different in-plane strains ($\varepsilon$).
In Fig. 7, we display the energy bands in the vicinity of the
Fermi level of the strained FM (LaPdO$_{3}$)$_2$ bilayer as an example.
Interestingly, under a compressive strain, the spin-down band 
moves upwards steadily as the strain strength increases while the spin-up
bands are gradually pushed downwards (see the left panels in Fig. 7).
At $\varepsilon = -1.0$ \%, the spin-down band is above the
Dirac point of the spin-up bands. In the meantime,
the spin-up Dirac point is now close to the Fermi level and 
remains so for the compressive strain ranging from -1.0 \% to -4.0 \%. 
At first glance, such movements of the energy bands caused by the compressive strain
seem to be similar to that due to the increased onsite Coulomb repulsion 
$U$ which enhances the intraatomic magnetization and hence the exchange
band splitting, as described in Sec. III.B. Nonetheless, a close look
suggests otherwise. For example, the calculated X spin magnetic moment
hardly changes with the in-plane strain (see Fig. 8). Instead, the size
of the mediating O spin magnetic moment increases significantly when
the strain is increased. For example, the O spin moment in the (LaPdO$_{3}$)$_2$ bilayer
with $U = 3.5$ eV for Pd is enhanced from -0.034 $\mu_B$ in the absence of
the strain to -0.068 $\mu_B$ at $\varepsilon = -4.0$ \%. Since the size 
of the O spin moment reflects the strength of the FM coupling
between the neighboring X atoms, as mentioned before, the compressive strain-enhanced
band spin spliting should result from the enhanced interatomic FM coupling 
caused by the stronger X $d$ and O $p$ orbital hybridization due to
the shortened X-O bondlengths in the strained (LaXO$_{3}$)$_2$ bilayers. 

When the SOC is included, these Dirac-like band crossing points
become gapped, as shown in the right panels in Fig. 7.
We can classify these local band gaps into two types, namely,
local gap 1 for the gap between the bands of the same spin
and local gap 2 for the gap between the bands of the opposite spins.
Interestingly, these two types of local band gaps can
occur simultaneously in the (LaPdO$_{3}$)$_2$ bilayer 
under various tensile strain amplitudes (Figs. 7 and 8)
However, local gap 2 disappears in the presence of 
the compressive strain, while local gap 1 survives all strains
although its energy position varies with the in-plane strain size.
Importantly, this demonstrates that the band gaps 
and related electronic properties of the (LaXO$_{3}$)$_2$ bilayers
can be significantly tuned by the in-plane strain.
\begin{figure}
\centering
\includegraphics[width=0.46\textwidth]{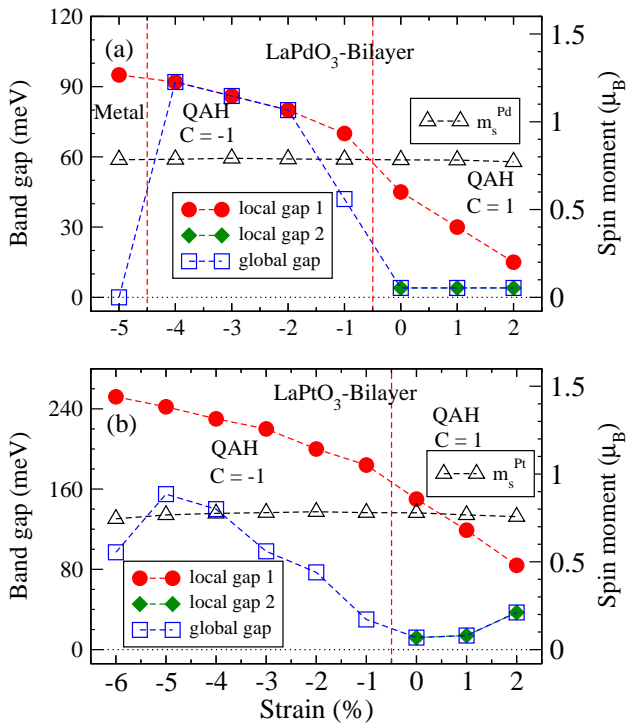}
\caption{\label{fig:Fig8} Calculated global gap,
type 1 local gap, type 2 local gap and also X spin moment ($m_s^X$)
in the FM (LaXO$_{3}$)$_{2}$ bilayers as a function of the biaxial
in-plane strain with $U = 3.5$ $(2.3)$ eV for Pd (Pt).
Red vertical dashed lines denote the phase boundares.}
\end{figure}

Let us now summarize the interesting effects of strains on
the properties of both (LaXO$_{3}$)$_{2}$ bilayers as 
the strain phase diagrams in Fig. 8.
Clearly, compressive in-plane strains increase local gap 1 
but suppress local gap 2. However, both local gaps 1 and 2 coexist
in the presence of tensile strains, although local gap 2 is
much smaller than local gap 1. Remarkably, both local gaps 
are topologically nontrivial with their Chern numbers having
opposite signs, namely, $C = -1$ for local gap 2 and $C = +1$ for local gap 1.
Meanwhile, the global band gap in the presence of tensile strains
is determined by local gap 2 while it originates from
local gap 1 in the presence of compressive strains.
Consequently, although both bilayers are QAH insulators 
under both types of strains, their topological phase changes
from $C = +1$ in the tensile strain regime to $C = +1$
in the compressive strain regime. This interesting finding suggests
that both (LaXO$_{3}$)$_{2}$ bilayers would offer a rich
playground for topological transport studies.  
Moreover, Fig. 8 shows that the global band gap can reach up 
to 92 meV in bilayer (LaPdO$_{3}$)$_{2}$ under a 4 \% in-plane 
compression and to 242 meV in bilayer (LaPtO$_{3}$)$_{2}$ 
when a 5$\%$ in-plane compression is applied. All these 
results thus indicate that tunable high temperature QAH 
phase could be realized in bilayers (LaXO$_{3}$)$_{2}$ 
by adjusting the in-plane strain.

Finally, we note that compared to the QAH phases predicted so far
in other real materials~\cite{weng}, the QAH phases in bilayer
(LaXO$_{3}$)$_{2}$ (X = Pd, Pt) have, at least, one distinct feature,
i.e., the toplogical band gaps can be opened not only in the crossing 
bands of the same spin but also in the crossing bands of opposite spins in the same
system. Moreover, the direction of the dissipationless edge current 
can be switched by either the onsite Coulomb repulsion $U$ or in-plane 
strain $\varepsilon$. This is similar to the cases considered previously~\cite{HongZhang2012},
in which the QAH gaps can be manipulated by the external electric
fields. Therefore, the QAHE predicted here in bilayers
(LaXO$_{3}$)$_{2}$ (X = Pd, Pt) would be superior for low-power
consumption nanoelectronic and spintronic applications.

\section{Conclusions}

In summary, we have performed a systematic first-principles DFT study of 
the magnetic and electronic properties of perovskite bilayers LaXO$_3$ (X = Pd, Pt) 
imbeded in the (111) (LaXO$_3$)$_{2}$/(LaAlO$_3$)$_{10}$ superlattices. 
Interestingly, we find that these TMO perovskite bilayers are high Curie 
temperature ferromagnets that would host both QAH and Dirac semimetal 
(see Fig. 9 in the Appendix) phases. Remarkably, in the QAH phase both 
the direction (the sign of Chern number) and spin-polarization of the 
chiral edge currents are tunable by either onsite electron correlation
or biaxial in-plane strain. Furthermore, the nontrivial band gap can
be enhanced up to 92 meV in the LaPdO$_3$ bilayer by the compressive 
in-plane strain, and can go up to as large as 242 meV when the Pd atoms 
are replaced by the heavier Pt atoms. By analyzing their underlying 
electronic band structures, we also uncover the microscopic mechanisms 
of the FM coupling and other interesting properties of the bilayers.
Our findings thus show that (111) perovskite LaXO$_3$ (X = Pd, Pt) 
(111) bilayers are quasi-2D high temperature FM insulators for investigating 
exotic quantum phases tunable by both onsite electron correlation
and biaxial in-plane strain, and also for advanced applications 
such as low-power nanoelectronics and oxide spintronics.

\begin{figure}
\centering
\includegraphics[width=0.46\textwidth]{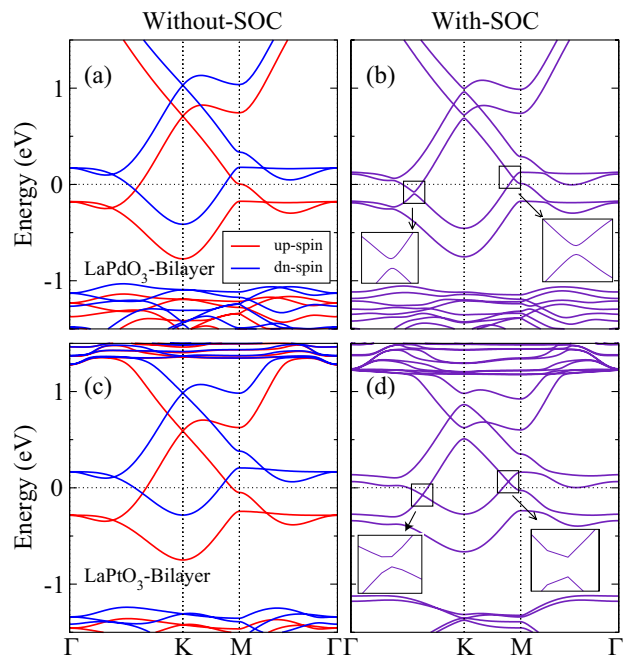}
\caption{\label{fig:Fig3} Band structures without (a,c) and with (b,d)
SOC of the (LaPdO$_{3}$)$_2$ (a,b) and (LaPtO$_{3}$)$_2$ (c,d) bilayers
from the GGA calculations. Zero refers to the
Fermi level. }
\end{figure}

\begin{acknowledgments}
The authors acknowledge the supports from the Ministry of Science and
Technology, National Center for Theoretical Sciences, and Academia
Sinica of the Republic of China. H. L. is also supported by the National Natural
Science Foundation of China under Grants No.11704046.
\end{acknowledgments}

\section*{APPENDIX: The GGA band structures}
The band structures of the (LaPdO$_{3}$)$_2$ and (LaPtO$_{3}$)$_2$ bilayers
from the GGA calculations are displayed in Fig. 9. As one can see from Figs. 9(a) and 9(c),
in the absence of SOC, there are two Dirac points near the Fermi level, one slightly below
the Fermi level along the $\Gamma$-K symmetry line and the other slightly above the Fermi level
along the K-M line. Therefore, both systems are a Dirac semimetal. When the SOC is included,
type 2 local gaps of 21 and 12 meV open at the Dirac points in bilayer (LaPdO$_{3}$)$_2$ 
[see Fig. 9(b)] and local gaps of 7 and 3 meV open in bilayer (LaPtO$_{3}$)$_2$ [Fig. 9(d)].
Nonetheless, both systems are still Dirac semimetals with their Fermi level cutting through 
the two Dirac cones below and above the Dirac points, respectively. 
Furthermore, type 1 local gaps of 36 [116] meV open at the Dirac points at the K point
significantly above the Fermi level in bilayer (LaPdO$_{3}$)$_2$ [(LaPtO$_{3}$)$_2$].

\end{document}